\documentclass[aps,prd,final,superscriptaddress,nofootinbib,nopacs,
nokeys,12pt,a4paper,twoside,notitlepage,preprintnumbers
]{revtex4-1}
\usepackage{amsmath}
\usepackage{amssymb}
\usepackage{mathtools}
\usepackage{graphicx}
\usepackage{color}

\usepackage[utf8]{inputenc}
\usepackage[T1]{fontenc}
\usepackage{lmodern}

\usepackage{mathrsfs}

\usepackage[hidelinks]{hyperref}

\usepackage{relsize,exscale}
\usepackage{tensor}

\usepackage[]{subdepth}
\usepackage{pgfplots}
\pdfoutput=1

\begin{document}
\title{The Thermal Feedback Effects on the Temperature Evolution during Reheating}

\author{Lei Ming}
\email[\tt]{minglei@smail.nju.edu.cn}
\affiliation{School of Physics, Nanjing University, Nanjing, 210093, China}

\begin{abstract}
	The time dependence of the temperature during the reheating process is studied. We consider the thermal feedback effects of the produced particles on the effective dissipation rate of the inflaton field, which can lead to enhanced production of particles. We parameterize the temperature dependence of the dissipation rate in terms of a Taylor expansion containing the vacuum decay rate and the thermal terms. By solving the Boltzmann equations for the energy densities of the inflaton and radiation, we provide analytic estimates for a general power-law dependence on the temperature. In this way we describe the entire reheating process. The maximum temperature of the reheating process and its dependence on model parameters are studied in different cases. The impact of the thermal feedback effects on the expansion history of the universe and the cosmic microwave background (CMB) is discussed. We also discuss the range of validity of our approach.
\end{abstract}


\maketitle
\tableofcontents

\section{Introduction}
Many properties of the observing universe today can be understood as the result of the processes happening during the early stage of our universe, e.g. the Big Bang Nucleosynthesis~\cite{Tanabashi:2018oca} and the abundance of the Standard Model matter~\cite{Davidson:2002qv, Fukugita:1986hr, Canetti:2014dka} as well as Dark Matter~\cite{Acharya:2008bk, Watson:2009hw, Acharya:2009zt}. In the  inflationary cosmology, reheating~\cite{Kofman:1997yn, Kofman:1994rk, Shtanov:1994ce} is a mechanism to set up the hot big bang initial conditions at the onset of the radiation-dominated era. After inflation~\cite{Starobinsky:1980te, Guth:1980zm, Linde:1981mu}, the scalar inflaton field decays and dissipates its energy into the produced particles via the interaction between them, and (re)heats the universe. The history--especially the thermal history--of the universe is strongly affected by the matter production dynamics after inflation, and the abundance of the constituents of the plasma is sensitive to the details of the thermal history~\cite{Bodeker:2020ghk}. Thus the knowledge of the temperature evolution of the early universe is crucial for cosmology. To understand in more detail about the thermal history of the early universe, we would like to investigate the temperature evolution during the reheating process.

The reheating process 
begins when the universe stops accelerating and ends at the moment when the energy density of the radiation is comparable to that of the expectation value $\varphi=\langle\phi\rangle$ of the scalar inflaton field $\phi$, $\rho_\gamma=\rho_\varphi$. It is well known that the particle production could be non-perturbative and in highly non-equilibrium with large couplings between the inflaton and other fields, and thus other mechanisms, such as the parametric resonance in preheating~\cite{Greene:1997fu, Felder:1998vq, Podolsky:2005bw}, need to be considered. In those models, the effective dissipation rate $\Gamma_\varphi$ that leads to the energy transfer from $\varphi$ to radiation may have a complicated time dependence, which makes it hard to study the evolution of the effective temperature $T$ as an effective parameter of the hot plasma. 

In this work, we assume that the plasma reaches equilibrium fast between the two zero crossings of the inflaton at the minimum of its potential, so that the occupation numbers can be expressed in terms of an effective temperature $T$~\cite{Harigaya:2013vwa, Davidson:2000er}.
If the reheating process starts from a cold universe with all the energy stored in the zero modes of the inflaton field, i.e. the initial value of $T$ and $\rho_\gamma$ are assumed to be zero, then after inflation the particle production process is dominated by the vacuum decay of the inflaton, which can be expressed by its temperature-independent decay rate $\Gamma_0$. As the temperature increases, the thermal feedback effects of the produced particles on $\Gamma_\varphi$, which are crucial and can lead to enhanced production of particles, should be considered. We will also discuss the situation where the reheating phase starts with non-zero initial temperature in Section~\ref{PRE}. In general, $\Gamma_\varphi$ has a complicated dependence on $T$ as shown in detailed studies~\cite{Drewes:2013iaa, Drewes:2015eoa, Mukaida:2012bz}, as different processes such as decays and scatterings can contribute to the dissipative rate and the phase space is temperature dependent due to the thermal correction to the effective masses. In general cases, we need to carefully study the particle production, thermalization, and resulting dissipation effect on the inflaton coherent oscillation.

The reheating temperature $T_R$ as the initial temperature of the radiation-dominated era is important for cosmology and is usually not the highest temperature during the early universe~\cite{Giudice:2000ex}. During reheating, the inflaton oscillates around the minimum of its effective potential and loses a small fraction of its energy per oscillation due to the relative smallness of $\Gamma\dot{\varphi}$. However, the total amount of energy transferred from $\varphi$ to radiation is the largest at early times because $\rho_\varphi$ is huge in the beginning and red-shifted at later times. In common cases, the universe becomes radiation-dominated when $\Gamma_\varphi=H$, with $H\equiv \dot{a}/a$ being the Hubble parameter in terms of the scale factor of the universe $a$, and shortly afterwards $\rho_\gamma$ exceeds $\rho_\varphi$. Since temperature evolution strongly affects the abundance of thermal relics which are sensitive to the thermal history, a quantitative understanding of the reheating process is necessary
.

We parameterize the effects as
\begin{equation}
    \Gamma_\varphi=\sum_{n=0}^{\infty}\Gamma_n\left(\frac{T}{m_\phi}\right)^n~,
\end{equation}
where $m_\phi$ is the inflaton mass and $n$ a positive integer. We can always Taylor expand $\Gamma_\varphi$ to be $\sum_{n=0}^{\infty}\Gamma_n(T/m_\phi)^n$ around a centain value of $T$. We study the set of scenarios where this can be done and the thermal part in $\Gamma_\varphi$ is dominated by one or few terms. In this way, it is possible to solve the Boltzmann equations for the energy densities $\rho_\varphi$ and $\rho_\gamma$ analytically, by considering the phases during which $\Gamma_0$ and $\Gamma_n$ dominate respectively and matching the solutions at the boundary. This piece-wise method approximates the complicated $\Gamma_\varphi$ to be a temperature-dependent monomial locally~\cite{Drewes:2014pfa, Drewes:2015coa} and has been used to obtain improved analytic estimates of the reheating temperature and maximal temperature in the early universe in the $n=2$ case. In~\cite{Co:2020xaf}, an arbitrary temperature dependence as well as a dependence of the dissipation rate on the scale factor are considered, and the increasing temperature is observed when the rate depends on the scale factor in some physical scenarios.

Our present paper generalize the results of~\cite{Drewes:2014pfa} and extend the research to the cases where the thermal term is a monomial of arbitrary $n$ powers. The $n=1$ case can appear in the corrections from the induced $1\rightarrow2$ decays. As will be seen, for a Bose enhanced $1\rightarrow2$ decay one generally has the form
\begin{equation}
    \Gamma_\varphi=\Gamma_0\left( 1+2f_B\left(\frac{m_\phi}{2}\right)\right)~,
\end{equation}
then we get a linear term in $T$ by expanding the Bose-Einstein distribution function $f_B$ with $m_\phi<T$.
The $n=2$ case appears quite generically in the high temperature regime when different processes such as $1\rightarrow3$ decays or $2\rightarrow2$ scatterings are included. For example in the $\alpha\phi\chi^3$ model one has~\cite{Drewes:2013iaa}
\begin{equation}
    \Gamma_\varphi\approx\frac{\alpha^2m_\phi}{3072\pi^3}+\frac{\alpha^2T^2}{768\pi m_\phi}~,
\end{equation}
Where $\alpha$ is the coupling constant between the inflaton and another scalar field $\chi$.
The higher power cases with $n>2$ can exist in some certain models~\cite{Garcia:2020wiy, Mukaida:2012bz}. 
In general, the interactions between inflaton and other fields in those models are higher-order operators in a non-renormalizable theory by dimensional arguments. 
Yet an analysis with arbitrary positive or even negative $n$ is meaningful especially in intermediate temperature regimes for the realistic models in which $\Gamma_\varphi$ is a complicated function of $T$. For example the thermal correction with $\Gamma_\varphi\propto T^n$  where $n=3,~4$ or $5$ may solve the cosmological moduli problem~\cite{Yokoyama:2006wt, Bodeker:2006ij}. Besides in curved spacetime with a finite temperature background, the effective dissipation rate might even contain negative $n$ powers functions of $T$~\cite{Ming:2019qty}. 

Our work generalizes the attempt to analytically solve the Boltzmann equations for the energy density of inflaton and radiation. 
The novel points of this paper are
\begin{enumerate}
	\item We describe the entire reheating process. 
	\item We present the dependence of the maximal temperature on model parameters.
	\item We discuss the impact of thermal effects on expansion history and the CMB.
	\item We discuss the range of validity for our approach.
\end{enumerate}
We will illustrate these points in the explicit models: the interaction between inflaton and other scalars or the gauge boson production from axion-like coupling. The details are discussed in Section~\ref{diss}.

In the following section, we will consider the case where $n$ is an arbitrary positive integer, and give analytic solutions to the Boltzmann equations for the energy densities of the inflaton and radiation, $\rho_\varphi$ and $\rho_\gamma$. The latter can determine the temperature evolution during reheating. With the analytic solution of the effective temperature $T$, we discuss 
the maximal temperature in different cases. We will also compare the analytic results to the numerical ones, showing that the piece-wise approximation is sufficient to catch the important characteristics of the reheating process. In Section~\ref{par}, we fix the model parameters in realistic models. Section~\ref{PRE} discuss the secenario in which there exists a preheating phase after inflation. The conclusions are presented in Section~\ref{conc}.  
\section{The solutions to the Boltzmann equations}\label{solu}
The Boltzmann equations for the energy densities $\rho_\varphi$ and $\rho_\gamma$, averaged over momentum and time across a few cycles of oscillations, are
\begin{eqnarray}
&&\frac{d\rho_\varphi}{dt}+3H\rho_\varphi+\Gamma_\varphi \rho_\varphi =0 \label{e22}\\
&&\frac{d\rho_\gamma}{dt}+4H\rho_\gamma-\Gamma_\varphi \rho_\varphi=0 \label{e23}
\end{eqnarray}
with 
\begin{equation}
    \rho_\gamma=\frac{\pi^2 g_*}{30}T^4~. \label{e24}
\end{equation}
Introducing the variables $\Phi\equiv\rho_\varphi a^3/m_\phi$, $R\equiv\rho_\gamma a^4$ and $x\equiv am_\phi$, 
the above equations can be rewritten as 
\begin{eqnarray}
    &&\frac{d\Phi}{dx}=-\frac{\Gamma_\varphi}{H x}\Phi \label{e25}\\
    &&\frac{d R}{dx}=\frac{\Gamma_\varphi}{H }\Phi \label{e26}
\end{eqnarray}
with
\begin{eqnarray}
    &&H=\left(\frac{8\pi}{3}\right)^{1/2}\frac{m_\phi^2}{M_P}\left(\frac{R}{x^4}+\frac{\Phi}{x^3}\right)^{1/2} \label{e27}
    \\
    &&T=\frac{m_\phi}{x}\left(\frac{30}{\pi^2g_*}R\right)^{1/4}~. \label{e28}
    \end{eqnarray}
Here $M_P$ and $g_*$ are the Planck mass and the number of degrees of freedom in the thermal bath, respectively.   
With these conventions, $x$ measures the dimensional scale factor $a$ in units of its value at the end of inflation, $x=a/a_{\rm end}$, and the initial value of $R$ is $R=0$ when $x=1$.

The reheating temperature $T_R$ is defined to be the temperature when reheating ends and $\rho_\varphi=\rho_\gamma$, i.e.
\begin{equation}
    \frac{R}{x^4}=\frac{\Phi}{x^3} \label{R=Phi}~.
\end{equation}
This usually coincides in good approximation with the moment when $\Gamma_\phi=H$, prior to which the fractional energy loss of $\varphi$ is negligible and one can in good approximation set $\Phi=\Phi_I$ as a constant and view it as an external source in~(\ref{e26}). If $\Gamma_\varphi$ is independent of $T$, i.e. $\Gamma_\varphi=\Gamma_0={\rm const}$, the usual estimated expression for $T_R$ can be found using $H=\sqrt{8\pi^3g_*/90}T^2/M_P$,
\begin{equation}
    T_R=\sqrt{\Gamma_0 M_P}\left(\frac{90}{8\pi^3g_*}\right)^{1/4}~.
    \label{TR}
\end{equation}

In this work, we consider the situation where one of the temperature dependent terms dominates in the relevant range of temperatures
\begin{equation}
\Gamma_\varphi=\Gamma_0+\Gamma_n\left(\frac{T}{m_\phi}\right)^n~.
\label{Gp}
\end{equation}
We use a local piece-wise approximation of $\Gamma_\varphi$ by monomials in $T$ when considering only one term at a time, and
the thermal behaviour of the early universe after inflation can be divided into two phases, during which $\Gamma_0$ and $\Gamma_n$ dominate respectively. A critical value of temperature at which the finite temperature effects start to dominate $\Gamma_\varphi$ can be defined as 
\begin{equation}
    T_n=m_\phi\left(\frac{\Gamma_0}{\Gamma_n}\right)^{1/n}~.
\end{equation}
Therefore the moment $x_n$ when $T=T_n$ can be viewed as the end of the $\Gamma_0$ phase, and the time-dependent temperature determined by $\Gamma_0$ at $x_n$ is the initial condition of the $\Gamma_n$ phase. 

The solutions for the $\Gamma_0$ phase, i.e. the case without thermal feedback, can be found as~\cite{Giudice:2000ex}
\begin{eqnarray}
    &&R=\frac{2}{5}A_0\left(x^{5/2}-1\right) \label{R0}\\
    &&T=m_\phi\left(\frac{2}{5}A_0\frac{30}{\pi^2 g_*}\right)^{1/4}\left(x^{-3/2}-x^{-4}\right)^{1/4}\label{T0}\\
    &&x_{\rm max}=\left(\frac{8}{3}\right)^{2/5} \\
    &&T_{\rm max}=T|_{x=x_{\rm max}}\approx0.6\left(\frac{\Gamma_0M_P}{g_*}\right)^{1/4}V_I^{1/8} \label{Tmax}~,
\end{eqnarray}
here $V_I$ is the value of the inflaton potential at the end of inflation, $x_{\rm max}$ is the moment when $T$ reaches its maximum $T_{\rm max}$, and $A_n$ is defined to be
\begin{equation}
    A_n=\frac{\Gamma_n M_P}{m_\phi^2}\sqrt{\Phi_I}\left(\frac{30}{\pi^2 g_*}\right)^{n/4}\left(\frac{3}{8\pi}\right)^{1/2}~.
    \label{An}
\end{equation}
A useful relation between $A_n$ and $A_0$ can be obtained as
\begin{equation}
    A_n=\frac{A_0\Gamma_n}{\Gamma_0}\left(\frac{30}{\pi^2g_*}\right)^{n/4}~.
\end{equation}

\begin{figure}[htbp]
    \centering
    \includegraphics[width=0.8\linewidth]{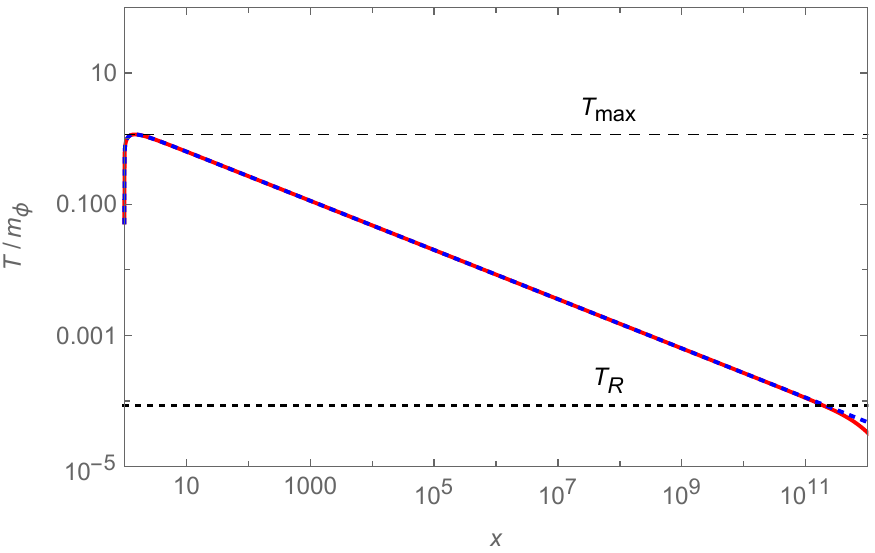}
    \caption{The time dependence of $T/m_\phi$ with the double logarithmic coordinates when $\Gamma_\varphi=\Gamma_0$. We show
    the comparison of the analytic solution (\ref{T0}) to a numerical solution of (\ref{e25})-(\ref{e28}). The numerical solution for (\ref{e26}) is in the red solid line and the analytic solution (\ref{T0}) in the blue dotted line respectively. The black dotted line is the value of $T_R$ in (\ref{TR}), and the black dashed line is the value of $T_{\rm max}$ in (\ref{Tmax}). The choices for the parameters are: $m_\phi=10^9~{\rm GeV}$, $\Gamma_0=10^{-8}~{\rm GeV}$, $\Phi_I=10^{20}$ and $g_*=100$.}
    \label{Tn=0}
\end{figure}

\begin{figure}[htbp]
\centering
    \includegraphics[width=0.8\linewidth]{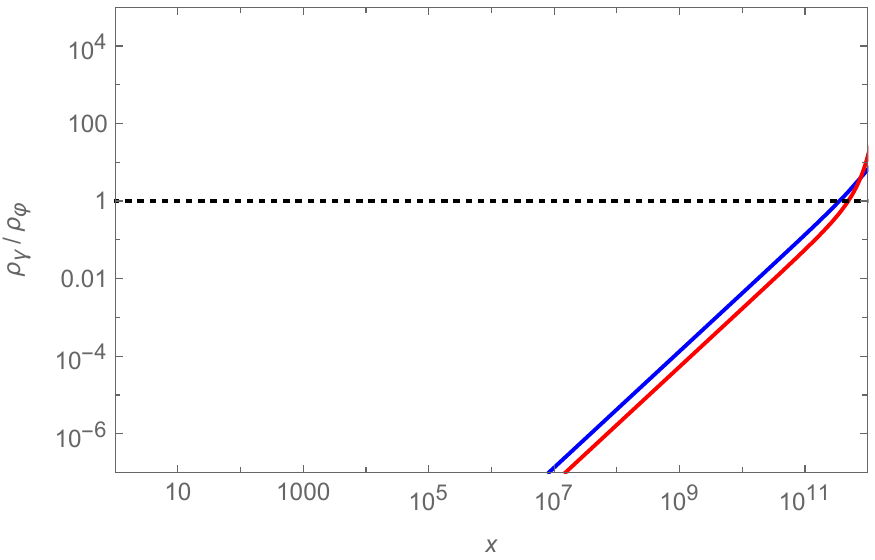}
    \caption{The time dependence of $\rho_\gamma/\rho_\varphi$ in the red line and $\Gamma_\varphi/H$ in the blue line with the double logarithmic coordinates when $\Gamma_\varphi=\Gamma_0$. After the moment $\rho_\gamma/\rho_\varphi=1$, the reheating ends and the universe becomes radiation dominated. It also coincides with the moment $\Gamma_\varphi=H$ and when (\ref{T0}) starts to fail. The choices for the parameters are: $m_\phi=10^9~{\rm GeV}$, $\Gamma_0=10^{-8}~{\rm GeV}$, $\Phi_I=10^{20}$ and $g_*=100$.} 
    \label{Hn=0}
\end{figure}

If $T_n<T_{\rm max}$, the temperature reaches $T_n$ when $x=x_n$ and $R=R_n$, where
\begin{eqnarray}
    &&x_n\approx1+\left(\frac{\Gamma_0}{\Gamma_n}\right)^{4/n}\frac{\pi^2 g_*}{30A_0}\label{x_i}~,\\
    &&R_n=\frac{2}{5}A_0(x_n^{5/2}-1)\approx\left(\frac{\Gamma_0}{\Gamma_n}\right)^{4/n}\frac{\pi^2 g_*}{30} \label{R_i}~.
\end{eqnarray}
Here we have expanded the solution (\ref{T0}) around $x=1$ since $1<x_n<x_{\rm max}\approx1.48$. Otherwise if $T_n>T_{\rm max}$ then the temperature can never reach the value where the $\Gamma_n$ term starts to dominate. This means the solution with $\Gamma_\varphi=\Gamma_0={\rm const}$ is valid, and the thermal feedback is negligible. The results for this case can be seen in Figure~\ref{Tn=0} and~\ref{Hn=0}. In Figure~\ref{Tn=0} we present the numerical solution for $T/m_\phi$ from (\ref{e26}) as the solid red line, and the analytic solution (\ref{T0}) as the dotted blue line, while the black dotted line represents the value of reheating temperature (\ref{TR}). The initial value of $T$ at the moment $x=1$ is set to be zero since we do not consider preheating here. With the numerical solution, Figure~\ref{Hn=0} shows the time dependence of $\rho_\gamma/\rho_\varphi$ and $\Gamma_\phi/H$ as the red and blue line respectively. It is clear that when $T_n>T_{\rm max}$, the reheating process is dominated by $\Gamma_0$ and (\ref{T0}) fully describes the time evolution of the effective temperature. The reheating process ends at the moment $\rho_\varphi=\rho_\gamma$, and this coincides well with $\Gamma_\phi=H$.

Below we will present the solutions for the $\Gamma_n$ dominated phase with the initial condition (\ref{R_i}) for arbitrary $n$, assuming that $T_n<T_{\rm max}$. During this phase, the equation (\ref{e26}) is 
\begin{equation}
    \frac{dR}{dx}=A_n R^{n/4}x^{3/2-n} \label{Req}~.
\end{equation}
\subsection{Case $n<4$}
In this case, the solution of the above equation is 
\begin{equation}
    R=\left(A_n\frac{1-n/4}{5/2-n}\left(x^{5/2-n}-x^{5/2-n}_n\right)+R^{1-n/4}_n\right)^{\frac{4}{4-n}}~,
    \label{me28}
\end{equation}
and therefore the solution for the temperature during the $\Gamma_n$ dominated phase is 
\begin{equation}
    T=m_\phi\left(\frac{30}{\pi^2g_*}\right)^{1/4}\frac{1}{x}\left(A_n\frac{1-n/4}{5/2-n}\left(x^{5/2-n}-x^{5/2-n}_n\right)+R^{1-n/4}_n\right)^{\frac{1}{4-n}}~.
    \label{me29}
\end{equation}
If $T_R<T_n<T_{\rm max}$, then the temperature $T_{\rm max}$ given by (\ref{Tmax}) is no longer applicable when $x>x_n$ since the $\Gamma_n$-term cannot be neglected after $T$ reaches $T_n$. The solution for all time will be a piecewise function given by (\ref{T0}) and (\ref{me29}), and the temperature reaches its maximum at the moment 
\begin{equation}\begin{aligned}
    \tilde{x}_{\rm max}&=\left(\frac{8-2n}{3}x^{5/2-n}_n+\frac{4(2n-5)}{3A_n}R^{1-n/4}_n\right)^{\frac{1}{5/2-n}}\\
    &\approx\left(\frac{8-2n}{3}\right)^{\frac{2}{5-2n}}x_n^\frac{n}{n-4}~.
    \label{me30}
\end{aligned}
\end{equation}
Considering that 
\begin{equation}\begin{aligned}
 &A_n\frac{1-n/4}{5/2-n}\left(x^{5/2-n}_{\rm max}-x^{5/2-n}_n\right)+R^{1-n/4}_n \\
 =&A_n\frac{1-n/4}{5/2-n}\left(\frac{8-2n}{3}x^{5/2-n}_n+\frac{4(2n-5)}{3A_n}R_n^{1-n/4}-x^{5/2-n}_n\right)+R_n^{1-n/4} \\
 =&\frac{2(1-n/4)}{3}A_nx^{5/2-n}_n+\frac{2(n-4)}{3}R^{1-n/4}_n+R^{1-n/4}_n \\
 =&\frac{2(1-n/4)}{3}A_nx_n^{5/2-n}+\frac{2n-5}{3}R^{1-n/4}_n \\
 =&\frac{A_n}{4}\tilde{x}_{\rm max}^{5/2-n}~,
 \end{aligned}
\end{equation}
one gets 
\begin{equation}\begin{aligned}
    \tilde{T}_{\rm max}&=m_\phi\left(\frac{30}{\pi^2g_*}\right)^{1/4}\tilde{x}^{-1}_{\rm max}\left(\frac{A_n}{4}\tilde{x}_{\rm max}^{5/2-n}\right)^{\frac{1}{4-n}} \\
    &=m_\phi\left(\frac{30}{\pi^2g_*}\right)^{1/4}\tilde{x}^{-\frac{3}{2(4-n)}}_{\rm max}\left(\frac{A_n}{4}\right)^{\frac{1}{4-n}} \\
    &\approx m_\phi\left(\frac{30}{\pi^2g_*}\right)^{1/4}\left(\frac{A_n}{4}\right)^{\frac{1}{4-n}}\left(\frac{8-2n}{3}\right)^{\frac{3}{(n-4)(5-2n)}}x_n^\frac{3n}{2(n-4)^2}
    ~.
    \label{me32}
    \end{aligned}
\end{equation}
In certain $n$ case, we can express $\tilde{T}_{\rm max}$ in terms of the model parameters \{$m_\phi$, $\Gamma_0$, $\Gamma_n$, $g_*$ and $\Phi_I$\} and $\Gamma_n$ further in terms of the inflaton coupling in realistic models. This will be discussed in more details in Section~\ref{par} and~\ref{diss}.

If $T_n<T_R<T_{\rm max}$, then the universe gets reheated very quickly due to the so called thermal resonance, and the radiation dominated era ($\rho_\gamma>\rho_\varphi$) starts in the $\Gamma_n$ regime
. In this situation, the maximum temperature and the reheating temperature are both roughly $\tilde{T}_{\rm max}$ because the universe reheats almost instantaneously at the moment $x=x_{\rm crit}$ when (\ref{R=Phi}) stands~\cite{Drewes:2014pfa}, i.e. $R=\Phi_I x$, and it corresponds to the solution of 
\begin{equation}
    \left(A_n\frac{1-n/4}{5/2-n}\left(x^{5/2-n}-x^{5/2-n}_n\right)+R^{1-n/4}_n\right)^{\frac{4}{4-n}}=\Phi_I x~.
    \label{eqx_cr}
\end{equation}
In the following we will focus on the cases where $T_R<T_n<T_{\rm max}$ and give the analytic solutions for $T$ as well as its maximum $\tilde{T}_{\rm max}$. 

More specifically, in the case $n<4$ we have $n=1,~2,~3$, below we present the results respectively. In the $n=1$ and $n=2$ cases, we will pin down the parameters by physical arguments and compare the analytic solution with the numerical one in realistic models.

\begin{itemize}
    \item $n=1$
    
The solutions (\ref{me28}), (\ref{me29}), (\ref{me30}) and (\ref{me32}) are
    \begin{eqnarray}
    &&R=\left(\frac{A_1}{2}\left(x^{3/2}-x^{3/2}_1\right)+R_1^{3/4}\right)^{4/3}\\
    &&T=m_\phi\left(\frac{30}{\pi^2g_*}\right)^{1/4}\frac{1}{x}\left(\frac{A_1}{2}\left(x^{3/2}-x^{3/2}_1\right)+R_1^{3/4}\right)^{1/3}\label{T1}\\
    &&\tilde{x}_{\rm max}=\left(2x_1^{3/2}-4\frac{R_1^{3/4}}{A_1}\right)^{2/3}\approx\left(\frac{4}{x_1}\right)^{1/3}\\
    &&\tilde{T}_{\rm max}\approx m_\phi\left(\frac{30}{\pi^2g_*}\right)^{1/4}\left(\frac{A_1}{8}\right)^{1/3}x_1^{1/6} \label{TM1}~,
    \end{eqnarray}
while the boundary condition is 
\begin{equation}
    T_1=m_\phi\frac{\Gamma_0}{\Gamma_1}~~~~{\rm at}~~~~x_1\approx1+\left(\frac{\Gamma_0}{\Gamma_1}\right)^4\frac{\pi^2g_*}{30A_0}.\label{Ti1}
\end{equation}

Below we present the comparison between the analytic solution and the numerical solution with the above set up. The temperature evolution can be seen in Figure~\ref{Tn=1}, and the evolution for $\rho_\gamma/\rho_\varphi$ and $\Gamma_\varphi/H$ in Figure~\ref{Hn=1}. At the beginning, the temperature is below $T_1$ and $\Gamma_\varphi$ is dominated by $\Gamma_0$, thus the evolution of $T$ is the same as that of the vacuum decays, (\ref{T0}). In the range $T_1<T<\tilde{T}_{\rm max}$, the dissipation can be described by the $\Gamma_1$ term, and the solution is (\ref{T1}) with the boundary conditions (\ref{Ti1}). It is clear that the $\Gamma_0$ term will dominate again after the temperature in the $\Gamma_1$ regime drops below $T_1$, and this extra $\Gamma_0$ phase holds until it reaches $T_R$. When $T_R<T_1<T_{\rm max}$ is fulfilled by the parameters, the piece-wise approximation for $\Gamma_\varphi$ works extremely well compared to the numerical solution. 
\begin{figure}[htbp]
    \centering
    \includegraphics[width=0.8\linewidth]{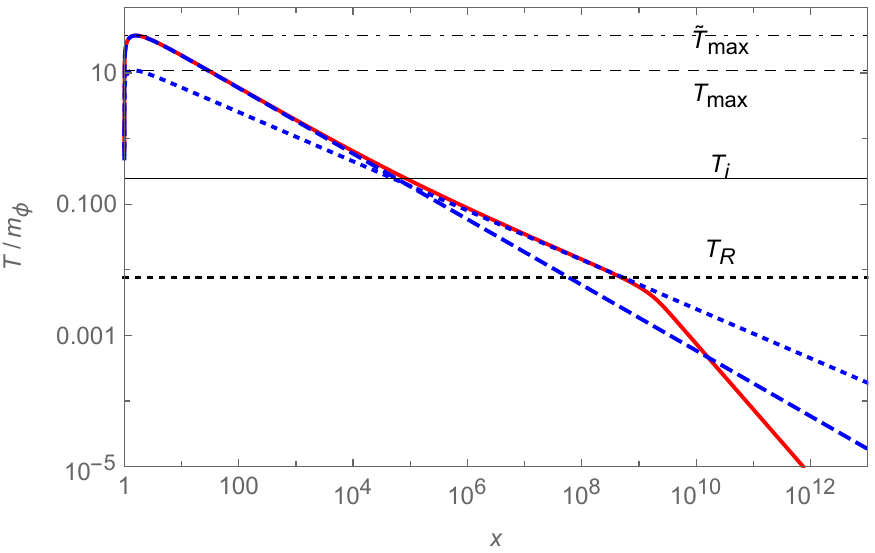}
    \caption{The time dependence of $T/m_\phi$ with the double logarithmic coordinates when $\Gamma_\varphi=\Gamma_0+\Gamma_1(T/m_\phi)$. The numerical solution for (\ref{e26}) is in the red solid line, the analytic solution (\ref{T0}) in the blue dotted line and (\ref{T1}) in the blue dashed line respectively. The black dotted line is the value of $T_R$ in (\ref{TR}), the black dashed line is the value of $T_{\rm max}$ in (\ref{Tmax}), the black solid line is the value of $T_1$ in (\ref{Ti1}), and the black dotted-dashed line is the value of $\tilde{T}_{\rm max}$ in (\ref{TM1}). The choices for the parameters are: $m_\phi=10^9~{\rm GeV}$, $\Gamma_0=8\times10^{-5}~{\rm GeV}$, $\Gamma_1=3.2\times10^{-4}~{\rm GeV}$, $\Phi_I=10^{20}$ and $g_*=100$. The choice for these values will be discussed within a realistic model in Section~\ref{par}.}
    \label{Tn=1}
\end{figure}
\begin{figure}[htbp]
    \centering
    \includegraphics[width=0.8\linewidth]{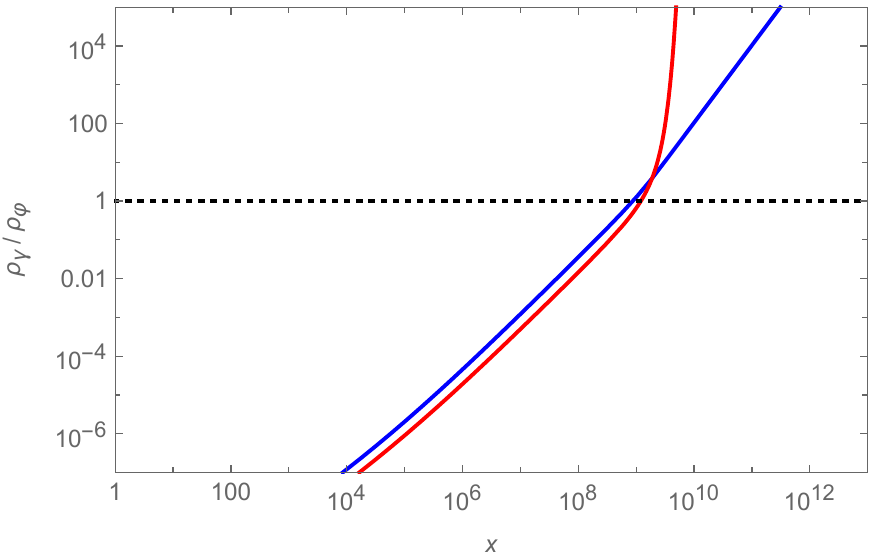}
    \caption{The time dependence of $\rho_\gamma/\rho_\varphi$ in the red line and $\Gamma_\varphi/H$ in the blue line when $\Gamma_\varphi=\Gamma_0+\Gamma_1(T/m_\phi)$. The moment $\rho_\gamma/\rho_\varphi=1$ happens when $\Gamma_\varphi=H$ and reheating ends. The choices for the parameters are the same with~\ref{Tn=1}: $m_\phi=10^9~{\rm GeV}$, $\Gamma_0=8\times10^{-5}~{\rm GeV}$, $\Gamma_1=3.2\times10^{-4}~{\rm GeV}$, $\Phi_I=10^{20}$ and $g_*=100$.}
    \label{Hn=1}
\end{figure}

\item $n=2$

The solutions (\ref{me28}), (\ref{me29}), (\ref{me30}) and (\ref{me32}) are
 \begin{eqnarray}
 &&R=\left(A_2\left(\sqrt{x}-\sqrt{x_2}\right)+\sqrt{R_2}\right) \\
 &&T=m_\phi\left(\frac{30}{\pi^2g_*}\right)^{1/4}\frac{1}{x}\sqrt{A_2\left(\sqrt{x}-\sqrt{x_2}\right)+R_2}\label{T2}\\
  &&\tilde{x}_{\rm max}=\frac{\left(4A_2\sqrt{x_2}-4\sqrt{R_2}\right)^2}{9A_2^2}\approx\frac{16}{9x_2}\\
 &&\tilde{T}_{\rm max}=m_\phi\left(\frac{30}{\pi^2g_*}\right)^{1/4}\left(\frac{3}{4}\right)^2\sqrt{\frac{A_2}{3}}x_2^{3/4}\label{TM2}~,
 \end{eqnarray} 
 while the boundary condition is 
 \begin{equation}
     T_2=m_\phi\sqrt{\frac{\Gamma_0}{\Gamma_2}}~~~~{\rm at}~~~~x_2\approx1+\left(\frac{\Gamma_0}{\Gamma_2}\right)^{2}\frac{\pi^2 g_*}{30A_0}.\label{Ti2}
 \end{equation}
Comparing our results of case $n=2$ with those in the section 2.2.4 of~\cite{Drewes:2014pfa}, we find that there exist some minor differences in this previous research for the expressions of $\tilde{x}_{\rm max}$, $\tilde{T}_{\rm max}$ and $x_2$. 

Again, we will extract the parameters in an illustrative model. 
and present the comparison between the analytic solution and the numerical solution.
The time evolution of the temperature and the ratio between energy densities are shown in Figure~\ref{Tn=2} and~\ref{Hn=2}. 

Figure~\ref{Tn=2} shows the time evolution of the effective temperature in three stages during reheating. At the beginning, the temperature is below $T_2$ and $\Gamma_\varphi$ is dominated by $\Gamma_0$, thus the evolution of $T$ is the same as that of the vacuum decays, (\ref{T0}). In the range $T_2<T<\tilde{T}_{\rm max}$, the dissipation can be described by the $\Gamma_2$ term, and the solution is (\ref{T2}) with the boundary conditions (\ref{Ti2}). The $\Gamma_0$ term will dominate again after the temperature drops below $T_2$ until the reheating ends at $T_R$.
 \begin{figure}[htbp]
    \centering
    \includegraphics[width=0.8\linewidth]{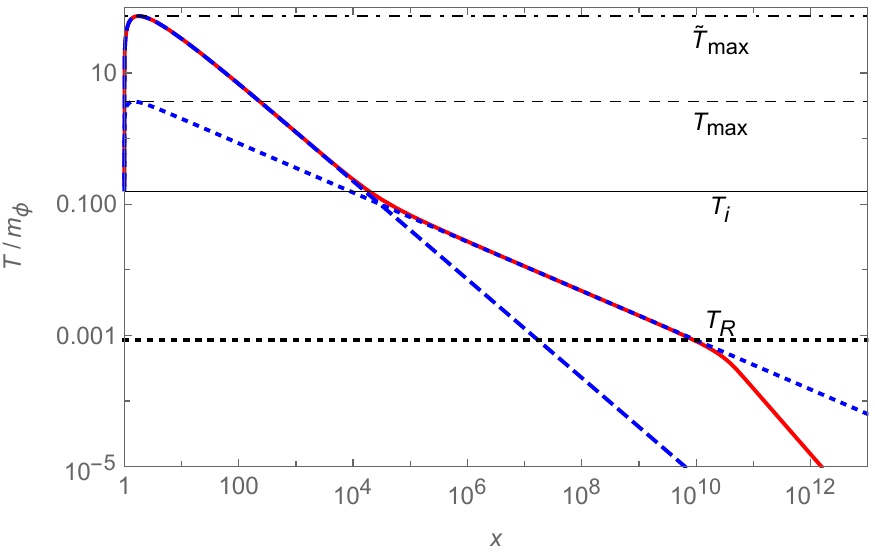}
    \caption{The time dependence of $T/m_\phi$ with the double logarithmic coordinates when $\Gamma_\varphi=\Gamma_0+\Gamma_2(T/m_\phi)^2$. The numerical solution for (\ref{e26}) is in the red solid line, the analytic solution (\ref{T0}) in the blue dotted line and (\ref{T2}) in the blue dashed line respectively. The black dotted line is the value of $T_R$ in (\ref{TR}), the black dashed line is the value of $T_{\rm max}$ in (\ref{Tmax}), the black solid line is the value of $T_2$ in (\ref{Ti2}), and the black dotted-dashed line is the value of $\tilde{T}_{\rm max}$ in (\ref{TM2}). The choices for the parameters are: $m_\phi=10^9~{\rm GeV}$, $\Gamma_0=10^{-6}~{\rm GeV}$, $\Gamma_2=4\times10^{-5}~{\rm GeV}$, $\Phi_I=10^{20}$ and $g_*=100$. The choice for these values will be discussed within a realistic model in Section~\ref{par}.}
    \label{Tn=2}
\end{figure}
 \begin{figure}[htbp]
    \centering
    \includegraphics[width=0.8\linewidth]{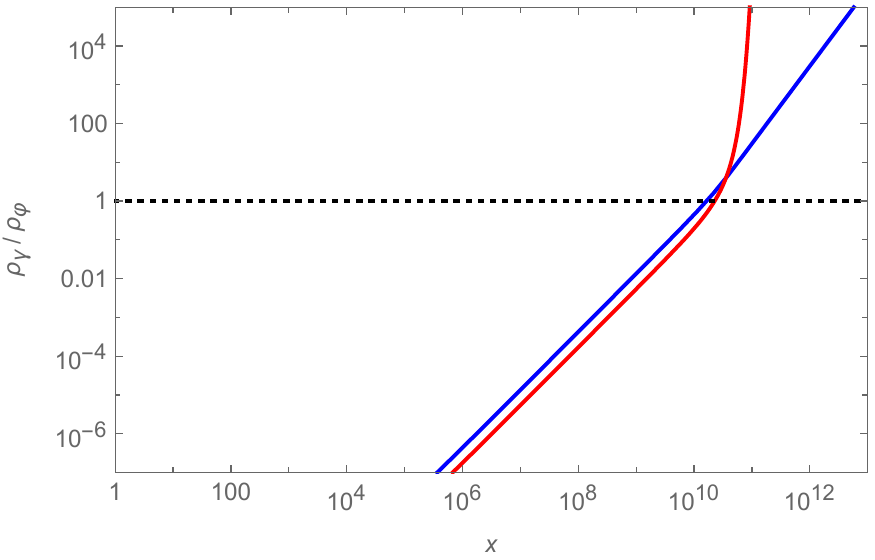}
    \caption{The time dependence of $\rho_\gamma/\rho_\varphi$ in the red line and $\Gamma_\varphi/H$ in the blue line when $\Gamma_\varphi=\Gamma_0+\Gamma_2(T/m_\phi)^2$. The moment $\rho_\gamma/\rho_\varphi=1$ happens when $\Gamma_\varphi=H$ and reheating ends. The choices for the parameters are the same with~\ref{Tn=2}: $m_\phi=10^9~{\rm GeV}$, $\Gamma_0=10^{-6}~{\rm GeV}$, $\Gamma_2=4\times10^{-5}~{\rm GeV}$, $\Phi_I=10^{20}$ and $g_*=100$.}
    \label{Hn=2}
\end{figure}

  \item $n=3$
  
The solutions (\ref{me28}), (\ref{me29}), (\ref{me30}) and (\ref{me32}) are
  \begin{eqnarray}
  &&R=\left(-\frac{A_3}{2}\left(\frac{1}{\sqrt{x}}-\frac{1}{\sqrt{x_3}}\right)+R_3^{1/4}\right)^4\\
  &&T=m_\phi\left(\frac{30}{\pi^2g_*}\right)^{1/4}\frac{1}{x}\left(\frac{A_3}{2\sqrt{x_3}}+R_3^{1/4}-\frac{A_3}{2\sqrt{x}}\right)\label{T3}\\
  &&\tilde{x}_{\rm max}=\left(\frac{2}{3\sqrt{x_3}}+\frac{4}{3A_3}R_3^{1/4}\right)^{-2}\approx\frac{9}{4x^{3}_3}\\
  &&\tilde{T}_{\rm max}=m_\phi\left(\frac{30}{\pi^2g_*}\right)^{1/4}\frac{A_3}{4}\left(\frac{2}{3}\right)^2x_3^{9/2}~,\label{TM3}
  \end{eqnarray}
  while the boundary condition is 
  \begin{equation}
      T_3=m_\phi\left(\frac{\Gamma_0}{\Gamma_3}\right)^{1/3}~~~~{\rm at}~~~~x_3=1+\left(\frac{\Gamma_0}{\Gamma_3}\right)^{4/3}\frac{\pi^2 g_*}{30A_0}~.\label{Ti3}
  \end{equation}
\end{itemize}

\subsection{Case $n=4$}
In this case, the equation (\ref{Req}) becomes 
    \begin{equation}
        \frac{dR}{dx}=A_4 R x^{-5/2}~,
    \end{equation}
    and its solution  is 
    \begin{equation}
        R=R_4 \exp\left(-\frac{2}{3}A_4\left(x^{-3/2}-x^{-3/2}_4\right)\right)~.\label{R4}
    \end{equation}
Therefore the time dependent function of the temperature is \begin{equation}
     T=m_\phi\left(\frac{30R_4}{\pi^2g_*}\right)^{1/4}\frac{1}{x}\exp\left(-\frac{1}{6}A_4\left(x^{-3/2}-x^{-3/2}_4\right)\right) \label{T4}~.
 \end{equation}
With the boundary condition
\begin{equation}
T_4=m_\phi\left(\frac{\Gamma_0}{\Gamma_4}\right)^{1/4}~~~~{\rm at}~~~~x_4=1+\frac{\Gamma_0}{\Gamma_4}\frac{\pi^2 g_*}{30A_0}~,\label{Ti4}   
\end{equation}
the maximal temperature can be found to be at the moment $\tilde{x}_{\rm max}=(A_4/4)^{2/3}$,
\begin{equation}
\tilde{T}_{\rm max}=m_\phi\left(\frac{30R_4}{\pi^2g_*}\right)^{1/4}\left(\frac{4}{A_4}\right)^{2/3}\exp\left(-\frac{2}{3}+\frac{A_4}{6}x_4^{-3/2}\right)~.
\label{TM4}
\end{equation}

\subsection{Case $n>4$}
 In this case, the solutions for $R$ and $T$ become 
\begin{eqnarray}
&R=\left|A_n\frac{1-n/4}{5/2-n}\left(x^{5/2-n}-x^{5/2-n}_n\right)+R^{1-n/4}_n\right|^{\frac{4}{4-n}}\label{R<4}  \\
&T=m_\phi\left(\frac{30}{\pi^2g_*}\right)^{1/4}\frac{1}{x}\left|A_n\frac{1-n/4}{5/2-n}\left(x^{5/2-n}-x^{5/2-n}_n\right)+R^{1-n/4}_n\right|^{\frac{1}{4-n}}~.\label{T5}
\end{eqnarray}
Here the reason for the absolute value function is that the variable inside may be negative, and we need to be careful when calculating the roots. The discussion for $n<4$ applies when the argument in the absolute value function is positive. However, one should notice that the above function of $T$ can have a pole in a certain range of parameters because the argument appears in the denominator due to $1/(4-n)<0$ and
\begin{equation}
    \begin{aligned}
    &R_n^{1-n/4}-\frac{1-n/4}{5/2-n}A_nx_n^{5/2-n} \\
    =&A_n\left(\frac{R_n^{1-n/4}}{A_n}-\frac{1-n/4}{5/2-n}x_n^{5/2-n}\right)\\
    \approx&A_n\left(\left(\frac{\Gamma_0}{\Gamma_n}\right)^{4/n-1}\left(\frac{\pi^2g_*}{30}\right)^{1-n/4}\frac{\Gamma_0}{A_0\Gamma_n}\left(\frac{\pi^2g_*}{30}\right)^{n/4}\right.\\
 \hspace{2cm}
    &\left.-\frac{1-n/4}{5/2-n}\left(1+\left(\frac{5}{2}-n\right)\left(\frac{\Gamma_0}{\Gamma_n}\right)^{4/n}\frac{\pi^2g_*}{30A_0}\right)\right)\\
    =&A_n\left(\frac{n/4-1}{5/2-n}+\frac{n}{4}\left(\frac{\Gamma_0}{\Gamma_n}\right)^{4/n}\frac{\pi^2g_*}{30A_0}\right)~.
    \end{aligned}
\end{equation}
The above expression will be negative when the second term in the bracket is smaller than $(n/4-1)/(n-5/2)$, while the latter goes from $0.1$ to $0.25$ in the range $n\geq5$. Thus in this situation, the denominator of (\ref{T5}) would be zero at some finite moment $x_0$. After the temperature reaches $T_n$, it will increase monotonically to infinity as $x\rightarrow x_0$. However, this solution was obtained under the assumption that the change in $\Phi_I$ is neglected, which is a good assumption until $\Gamma_\varphi= H$. This would break down before the temperature reaches the divergence simply because of energy conservation. The temperature must remain lower than the temperature that one would get if the inflaton would instantaneously transfer all its energy into radiation. 
This means that reheating would end soon after $x_n$ at some moment $x_{\rm crti}$, the solution of (\ref{eqx_cr}), since the energy of inflaton field is finite and thus the dissipation ends before $x_0$. This corresponds to $T_n<T_R$. In most cases, there does not exist an analytic solution for $x_{\rm crit}$, and the maximal temperature and the reheating temperature are both given by (\ref{T5}) at $x_{\rm crti}$. 

\section{Fixing the parameters in realistic models}\label{par}
In the previous section, we have given the analytic solutions as well as the comparison with the numerical results. Below we will first discuss how to determine the parameters in realistic models for the presented figures. 

For the case $n=1$, let us now consider in a model in which the perturbative reheating is dominated by the axion-like coupling between the scalar inflaton field and the gauge boson. The interaction term in the Lagrangian has the form
\begin{equation}
	\mathcal{L}_{\rm int}=\frac{\alpha}{\Lambda}\phi F_{\mu\nu}\tilde{F}^{\mu\nu}~,
\end{equation}
in which $F_{\mu\nu}$ is the field strength tensor of the vector bosons, $\alpha$ a dimensionless number and $\Lambda$ a mass scale. In the range where the inflaton mass is larger than the mass of the bosons and both are smaller than the temperature, the effective dissipation rate can be approximated as~\cite{Carenza:2019vzg}
\begin{equation}
	\Gamma_\varphi\approx\frac{\alpha^2}{4\pi}\frac{m_\phi^3}{\Lambda^2}\left(1+2f_B\left(\frac{m_\phi}{2}\right)\right)~.
\end{equation}
The Bose-Einstein distribution $f_B(y)$ can be expanded with $y<T$ to be $T/y$, and we get a linear term in temperature for $\Gamma_\varphi$,
\begin{equation}
	\Gamma_\varphi\approx\frac{\alpha^2}{4\pi}\frac{m_\phi^3}{\Lambda^2}\left(1+\frac{4T}{m_\phi}\right)~.
\end{equation}
We set the mass scale $\Lambda$ to be Planck mass $M_p$, the largest scale in our investigation. Introducing the dimensionless coupling  $\tilde{\alpha}\equiv\alpha m_\phi/M_P$, we would like $\tilde{\alpha}$ to be neither too large to trigger the parametric resonance~\cite{Drewes:2019rxn}, nor too tiny to be unrealistic. In the numerical simulation we set $\tilde{\alpha}=10^{-6}$, $m_\phi=10^9~{\rm GeV}$ and $g_*=100$. Then in the above notation we have
\begin{equation}
	\Gamma_0=\frac{1}{4}\Gamma_1=\frac{\tilde{\alpha}^2m_\phi}{4\pi}\approx8\times10^{-5}~{\rm GeV}.\label{G1A}
\end{equation}
For the estimation of the initial value of $\Phi_I$, considering that at the beginning of reheating $x=1$, the value of the  inflaton potential can be expressed as
\begin{equation}
	V_I\approx\frac{3}{4}\rho_\varphi=\frac{3}{4}\rho_\varphi x^3=\frac{3}{4}\rho_\varphi a^3m_\varphi^3=\frac{3}{4}\Phi_I m_\phi^4~.\label{VI}
\end{equation}
If the oscillation happens in the quadratic regime, $V_I\sim m_\phi^2\varphi_{\rm end}^2/2$, together with the fact that the field value at the end of inflation $\varphi_{\rm end}$ is usually near the Planck scale,
one has
\begin{equation}
	\Phi_I=\frac{4V_I}{3m_\phi^4}\approx\frac{2m_\phi^2\varphi_{\rm end}^2}{3m_\phi^4}\sim\frac{M_P^2}{m_\phi^2}~.\label{PI}
\end{equation}
Thus we can estimate that $\Phi_I=10^{20}$.

For the $n=2$ case, we consider that the perturbative reheating is dominated by the interaction
\begin{equation}
	\mathcal{L}_{\rm int}=\frac{1}{3!}\alpha\phi\chi^3
\end{equation} between the inflaton and another scalar field $\chi$, where $\alpha$ is the dimensionless coupling constant. In the limit of vanishing $\chi$-quasiparticle width and with $m_\chi\ll m_\phi$, the effective dissipation rate can be approximated as~\cite{Drewes:2013iaa}
\begin{equation}
	\Gamma_\varphi\approx\frac{\alpha^2m_\phi}{3072\pi^3}+\frac{\alpha^2T^2}{768\pi m_\phi}~.  
\end{equation}
Thus in our notation we have
\begin{equation}
	\Gamma_2=4\pi^2\Gamma_0=\frac{\alpha^2 m_\phi}{768\pi}~.
\end{equation}
We set the parameters to be
$\alpha=10^{-5}$, $m_\phi=10^9~{\rm GeV}$, $\Phi_I=10^{20}$ and $g_*=100$. In this set up,
\begin{equation}
	\Gamma_0\approx10^{-6}~{\rm GeV}~~,~~\Gamma_2\approx4\times10^{-5}~{\rm GeV}~.
\end{equation}
\section{The scenarios with the existence of preheating}\label{PRE}
In Section~\ref{solu}, we solve the Boltzmann equations for $\rho_\varphi$ and $\rho_\gamma$ in cases where $n$ is arbitrary. The initial value of $R$ (and thus $T$) at the beginning of reheating is set to be zero since we are only concerned with the perturbative reheating and the thermal feedback effects of the produced particles. However, the parametric resonance generally exists
with relatively large couplings between the inflaton and other fields. With the resonant matter production in the short preheating phase before the reheating process, the initial conditions need to be modified. Assuming that the inflaton dissipates parts of its energy into other degrees in preheating, quantified as 
\begin{equation}
    \rho_\gamma=\epsilon\rho_\varphi~,
\end{equation}
where $0\leq\epsilon\leq1$. If the parametric resonance is extremely efficient then $\epsilon=1$. On the other hand, $\epsilon=0$ corresponding to our previous assumption. Considering that the duration of preheating is usually much shorter than reheating, the initial conditions of the reheating process can be written as (the estimation of $\Phi_I$ still follows (\ref{PI}) at the end of inflation)
\begin{equation}
    R_I=\rho_\gamma a^4=\epsilon\rho_\varphi a^4=\epsilon\Phi_I a m_\phi\simeq\epsilon\Phi_I~,
\end{equation}
and 
\begin{equation}
    \Phi'_I=(1-\epsilon)\rho_\varphi \frac{a^3}{m_\phi}=(1-\epsilon)\Phi_I~.
\end{equation}
In this way the effective temperature at the beginning of reheating is then
\begin{equation}
    T_I=\left(\frac{30}{\pi^2g_*}\rho_\gamma\right)^{1/4}=\left(\frac{30}{\pi^2g_*}\epsilon\Phi_I m_\phi^4\right)^{1/4}~.
\end{equation}
Adopting the polynomial approximation of $\Gamma_\varphi$, one would find that the thermal term $\Gamma_n$ could dominate at the beginning of the reheating process, if $T_I$ is larger than $T_n$ in a certain $n$ case. And this corresponds to
\begin{equation}
    \frac{30}{\pi^2g_*}\epsilon\Phi_I >\left( \frac{\Gamma_0}{\Gamma_n}\right)^{4/n}~.
\end{equation}
Therefore the evolution of the effective temperature will be modified with the initial conditions determined by $\epsilon$ and the values of $\Gamma_n$ in practical models and the method used in the last section still applies.

\section{Discussions}\label{diss}
Our work generalizes the attempt to analytically solve the Boltzmann equations for the energy density of inflaton and radiation. 
The novel points in this paper are
\begin{enumerate}
	\item We describe the entire reheating process. 
	\item We present the dependence of the maximal temperature on model parameters.
	\item We discuss the impact of thermal effects on expansion history and the CMB.
	\item We discuss the range of validity for our approach.
\end{enumerate}
We have illustrated these points in the following explicit models: the interaction between inflaton and other scalars or the gauge boson production from axion-like coupling. Below we will discuss them in details.

\subsection{The description of the entire reheating process}
To solve the entire reheating process we need to locally use the power law approximation and match the solutions, as done in Section~\ref{solu} by the piece-wise approximation. In general, a $\Gamma_0$ piece is needed otherwise it is impossible to reheat the universe unless one assumes the presence of an intial radiation bath from preheating. Besides, the piece-wise approximation can be extended to the cases where $\Gamma_\varphi$ has a complicated dependence on $T$ other than a power law, and we can expand it to contain more terms and match the solutions piece by piece. 

In the situation where the temperature starts from $T=0$, we present the proper matching of the piece-wise solutions. For example, in the case $n=1$,
Figure~\ref{Tn=1} shows the time evolution of the effective temperature in three stages during reheating. At the beginning, the temperature is below $T_1$ and $\Gamma_\varphi$ is dominated by $\Gamma_0$, thus the evolution of $T$ is the same as that of the vacuum decays, (\ref{T0}). In the range $T_1<T<\tilde{T}_{\rm max}$, the dissipation can be described by the $\Gamma_1$ term, and the solution is (\ref{T1}) with the boundary conditions (\ref{Ti1}). The $\Gamma_0$ term will dominate again after the temperature drops below $T_1$ until the reheating ends at $T_R$. Figure~\ref{Hn=1} also indicates the locally power law approximation works well during the entire reheating process.

\subsection{The dependence of the maximal temperature on model parameters}
Our analytic results permits to discuss explicitly the dependence of the maximal temperature on model parameters. Taking (\ref{TM1}) as an example, in the case $n=1$ the dependence of $\tilde{T}_{\rm max}$  on the model parameters \{$m_\phi$, $\Gamma_0$, $\Gamma_1$, $g_*$ and $\Phi_I$\} can be obtained explicitly by (\ref{An}) and (\ref{Ti1}),
\begin{equation}
	\begin{aligned}
	\tilde{T}_{\rm max}&\approx m_\phi\left(\frac{30}{\pi^2g_*}\right)^{1/4}\left(\frac{A_1}{8}\right)^{1/3}x_1^{1/6}\\
	&=\frac{m_\phi}{2}\left(\frac{30}{\pi^2g_*}\right)^{1/3}\frac{\Gamma_1^{1/3}M_P^{1/3}\Phi_I^{1/6}}{m_\phi^{2/3}}\left(\frac{3}{8\pi}\right)^{1/6}\left(1+\left(\frac{\Gamma_0}{\Gamma_1}\right)^4\frac{\pi^2g_*}{30A_0}\right)^{1/6}\\
	&\approx\left(\frac{3}{8\pi}\right)^{1/6}\left(\frac{30}{\pi^2}\right)^{1/3}\frac{m_\phi^{1/3}\Gamma_1^{1/3}M_P^{1/3}\Phi_I^{1/6}}{2g_*^{1/3}} ~,\label{dep}
	\end{aligned}
\end{equation}
the last line is calculated by ignoring the second term in the bracket above, 
considering that $1<x_1<x_{\rm max}\approx1.48$ and thus $x_1^{1/6}\approx1$. Together with (\ref{G1A}) and (\ref{VI}), we have in the axion-like model
\begin{equation}
	\tilde{T}_{\rm max}\propto \alpha^{2/3} m_\phi^{2/3}V_I^{1/6}g_*^{-1/3}~.
\end{equation}
It is clear that the maximal temperature increases with the axion-like coupling, the inflaton mass and the value of inflaton potential at the end of inflation, and decreases with the number of degrees of freedom in the bath. If more accurate dependence is required, one can take the second term in the bracket of (\ref{dep}) into account and get another term depending on $\Gamma_0$, which further depends on $\alpha$.

\subsection{The impact of thermal effects on expansion history and the CMB}
The thermal feedback of the produced particles has impact not only on the thermal history of the universe, but also on the expansion history. The expansion history can be indeed affected significantly for $n<2$ and the reheating ends much earlier than expected. To see this, we consider that the Hubble parameter $H$ is larger than the dissipation rate $\Gamma_\varphi$ at the moment $\tilde{x}_{\rm max}$ when the temperature reaches its maximum. In the case $n\ge2$, $H$ approximately scales as $\propto T^2$ near the end of reheating and $\Gamma_\varphi$ scales as $\propto T^n$ during the $\Gamma_n$ dominated stage. Thus $\Gamma_\varphi$ decreases faster than $H$ with a decreasing $T$, and $\Gamma_\varphi=H$ happens during the $\Gamma_0$ dominated stage after $T$ drops below $T_n$, as shown in Section~\ref{solu}. However, in the $n=1$ case, $\Gamma_\varphi$ decreases slower than $H$ and it is probable that they meet each other in the $\Gamma_1$ dominated stage. In this situation, the duration of reheating is significantly shortened and the reheating temperature has to be modified, corresponding to $T_1<T_R<\tilde{T}_{\rm max}$.

We can further find the constraints on $\Gamma_1$ using the solution (\ref{T1}) of $T$ during the $\Gamma_1$ stage. Together with (\ref{e27}) and that $\Gamma_\varphi\simeq\Gamma_1 T/m_\phi$, the reheating temperature $T_R$ is obtained at the moment $\Gamma_\varphi=H$,
\begin{equation}
	T_R=\frac{\delta+\sqrt{\delta^2+4\gamma^3A_1}}{2\gamma^2}~,
\end{equation}
where
\begin{eqnarray}
	&&\delta\equiv\frac{4A_1}{m_\phi^3}\left(\frac{\pi^2g_*}{30}\right)^{3/4}\left(-\frac{1}{2}+\frac{1}{4}\left(\frac{\Gamma_0}{\Gamma_1}\right)^4\frac{\pi^2g_*}{30A_0}\right)\\
	&&\gamma\equiv2\sqrt{\frac{8\pi}{3}}\frac{\sqrt{\Phi_I}}{\Gamma_1M_P}\left(\frac{\pi^2g_*}{30}\right)^{3/4}~.
\end{eqnarray}
Considering that $T_1\equiv m_\phi\Gamma_0/\Gamma_1$, the condition $T_1<T_R$ then gives $\Gamma_1>\Gamma_{\rm 1,crit}$ with
\begin{equation}
	\begin{aligned}
	\Gamma_{1,{\rm crit}}&=\left(\frac{32\pi\Phi_Im_\phi^2\Gamma_0^2}{3M_P^2}-\frac{\pi^2g_*\Gamma_0^4}{30m_\phi^2}\right)^{1/4}\left(\frac{2\Phi_I}{m_\phi^2}-\frac{2\Gamma_0M_P\sqrt{\Phi_I}}{m_\phi^4}\sqrt{\frac{3}{8\pi}}\right)^{-1/4}\\
	&\approx\left(\frac{16\pi}{3}\right)^{1/4}m_\phi\sqrt{\frac{\Gamma_0}{M_P}}~.
	\end{aligned}
\end{equation}
The last step uses $\Gamma_0\ll m_\phi$ and $\Phi_I\sim M^2_P/m_\phi^2$ to drop the second terms in the two brackets. 

The reheating phase affects the cosmic microwave background (CMB) via its effects on the expansion history of the universe, since the equation of state during reheating is different from the one in inflationary or radiation dominated era, and the expansion history affect the way how physical scales at present time and during inflation relate to each other. 
While we can only observe the time-integrated effect, the quantities that the CMB is directly sensitive to are the duration of the reheating era in terms of the e-folds $N_{\rm re}$ and the averaged equation of state $\bar{\omega}_{\rm re}$ during reheating, with~\cite{Drewes:2019rxn} 
\begin{equation}
	\bar{\omega}_{\rm re}=\frac{1}{ N_{\rm re}}\int_0^{ N_{\rm re}}\omega(N)dN~.
\end{equation}
In the case $n=1$, when $\Gamma_1$ is so small that $T_1>T_{\rm max}$, the temperature evolution will just be the one of perturbative decay, a $\Gamma_0$ dominated stage. With $\Gamma_1$ being in a intermediate range, the reheating process contains several stages as decsribed. However, if $\Gamma_1>\Gamma_{1,{\rm crit}}$, the duration of reheating $N_{\rm re}$ will be significantly shortened, the thermal feedback effects has impact on not only the thermal history but also the expansion history, and therefore affects the predication of the observable CMB.

\subsection{The discussion of the range of validity}
We now discuss under what circumstances can our approach be applied. In the $n=1$ case, if we assume the perturbative reheating is dominated by the axion-like coupling, then by (\ref{G1A}) we have
\begin{equation}
	T_1\equiv m_\phi\frac{\Gamma_0}{\Gamma_1}=\frac{m_\phi}{4}~,
\end{equation}
while
\begin{equation}
\Gamma_0=\frac{1}{4}\Gamma_1=\frac{\alpha^2m_\phi^3}{4\pi M_P^2}~.
\end{equation}
To avoid the parametric resonance, the coupling $\alpha$ needs to be relatively small. In the Section 3.4 of~\cite{Drewes:2019rxn}, this upper bound can be found as
\begin{equation}
\tilde{\alpha}\equiv\alpha\frac{m_\phi}{\Lambda}\ll\min\left(\frac{m_\phi}{4\varphi_{\rm end}},~\frac{m_\phi}{5\sqrt{\varphi_{\rm end}M_P}}\right)~.
\end{equation}
If $\Lambda$ and $\varphi_{\rm end}$ are both set to be $M_P$, then one gets 
\begin{equation}
	\alpha\ll0.2~.
\end{equation}
On the other hand, with $V_I\sim m_\phi^2 M_P^2/2$ the $T_{\rm max}$ in (\ref{Tmax}) can be expressed as 
\begin{equation}
T_{\rm max}\simeq0.6\frac{\alpha^{1/2}m_\phi^{3/4}}{\left(4\pi\right)^{1/4}M_P^{1/2}}\frac{M_P^{1/4}}{g_*^{1/4}}m_\phi^{1/4}M_P^{1/4}=\frac{0.6m_\phi\sqrt{\alpha}}{\left(4\pi\right)^{1/4}g_*^{1/4}}~,
\end{equation} 
then the condition $T_1<T_{\rm max}$ translates into 
\begin{equation}
\alpha>\left(\frac{(4\pi)^{1/4}g_*^{1/4}}{4*0.6}\right)^{2}\simeq0.58\sqrt{g_*}~.
\end{equation}
These two conditions are contradictory with $g_*\ge1$, concluding that the thermal effects cannot modify the thermal history without triggering a parametric resonance. However, this conclusion relies on the assumptions that $\Lambda=M_P$ and $\varphi_{\rm end}=M_P$. While the former is reasonable, the latter may not be true in small field models in which the inflation ends at a sub-Planckian field value. Besides, if the perturbative reheating is dominated by more than one kind of interactions, the relation between $\Gamma_0$ and $\Gamma_1$ is usually not simply linear. For example if there exist a large Yukawa coupling and a small axion coupling, then the main contribution for $\Gamma_0$ comes from the Yukawa interaction and $\Gamma_1$ from the axion-like coupling since the former is not Bose enhanced. Thus we conclude that the thermal effects can only modify the thermal history if $\varphi_{\rm end}$ is not too large, if one has perturbative reheating in mind. The stepwise evolution of $T$ presented in the above figures still holds schematically, at least within the EFT framework. All in all, these does not inherently invalidate the method used, as long as the non-perturbative dissipation rate of the reheating process can locally be written in the form $\Gamma_n(T/m_\phi)^n$.

\section{Conclusions}\label{conc}
The reheating phase, which populates the universe with hot plasma and sets up the initial conditions for the radiation dominated era, affects the expansion and thermal history of the early universe and may produce relics such as gravitational waves, baryon asymmetry or dark matter. The information of the reheating process, especially the maximum effective temperature during reheating and the reheating temperature at the beginning of the radiation era, is crucial to understand how our universe evolves and how the matters are produced. 

In this work, we study the time dependence of the temperature during the reheating process.
The effective dissipation rate $\Gamma_\varphi$ of the inflaton is assumed to be a polynomial containing the vacuum decay of the inflaton represented by $\Gamma_0$, as well as a thermal term $\Gamma_n (T/m_\phi)^n$ which is an arbitrary power monomial of the effective temperature. By the piece-wise approximation of $\Gamma_\varphi$, we then solve the Boltzmann equations for the energy densities of inflaton and radiation, $\rho_\varphi$ and $\rho_\gamma$, and give analytic solutions for the time-dependent temperature. In this way, we describe the thermal history of the reheating process. The maximum temperature in the universe and its dependence on model parameters
are discussed. We also compare the analytic results to the numerical ones, which shows that the piece-wise approximation method is powerful and sufficient for us to study the information of reheating process.

Our work generalizes previous research in which the case $n=2$ was studied. We solve the equations governing the evolution of the energy densities of inflaton and radiation in arbitrary $n$ case, and we give practical examples in which the numerical results match the analytic ones well when $n=1$ or $2$. We find that some expressions in~\cite{Drewes:2014pfa} need to be modified in the $n=2$ case. We also study the situation where there exists a preheating phase after inflation: it changes the initial conditions of reheating for the effective temperature, and therefore the reheating could be dominated by the thermal feedback effects from the beginning. The impact of thermal feedback effects on the expansion history of the universe and therefore on the predication of CMB is discussed.

The result can contribute to the knowledge of the early epochs of cosmic history, and also its thermal relics including dark matter. For example, a large plasma temperature allows the production of heavier particles from inflaton, while this is forbidden via perturbative decays for kinematic reasons. Besides, the abundance of relics that is out of equilibrium before freeze-out is sensitive to the thermal history.

\vspace{6pt} 




\acknowledgments
This research was funded in parts by the NSF China (11775110, 11690034), the European Union’s Horizon 2020 research and innovation programme (RISE) under the Marie Sklodowska-Curie grant agreement (644121), and the Priority Academic Program Development for Jiangsu Higher Education Institutions (PAPD). The author would like to thank Marco Drewes, Edna Cheung and Hui Xu for their helpful comments on the manuscript.







\begin{thebibliography}{999}
\bibitem{Tanabashi:2018oca}
M.~Tanabashi \textit{et al.} [Particle Data Group],
``Review of Particle Physics,''
Phys. Rev. D \textbf{98} (2018) no.3, 030001
\href{https://journals.aps.org/prd/abstract/10.1103/PhysRevD.98.030001}{doi:10.1103/PhysRevD.98.030001}

\bibitem{Fukugita:1986hr}
M.~Fukugita and T.~Yanagida,
``Baryogenesis Without Grand Unification,''
Phys. Lett. B \textbf{174} (1986), 45-47
\href{https://www.sciencedirect.com/science/article/abs/pii/0370269386911263?via%3Dihub}{doi:10.1016/0370-2693(86)91126-3}

\bibitem{Canetti:2014dka}
L.~Canetti, M.~Drewes and B.~Garbrecht,
``Probing leptogenesis with GeV-scale sterile neutrinos at LHCb and Belle II,''
Phys. Rev. D \textbf{90} (2014) no.12, 125005
\href{https://journals.aps.org/prd/abstract/10.1103/PhysRevD.90.125005}{doi:10.1103/PhysRevD.90.125005}
\href{https://arxiv.org/abs/1404.7114}{[arXiv:1404.7114 [hep-ph]]}.

\bibitem{Davidson:2002qv}
S.~Davidson and A.~Ibarra,
``A Lower bound on the right-handed neutrino mass from leptogenesis,''
Phys. Lett. B \textbf{535} (2002), 25-32
\href{https://www.sciencedirect.com/science/article/pii/S0370269302017355?via%3Dihub}{doi:10.1016/S0370-2693(02)01735-5}
\href{https://arxiv.org/abs/hep-ph/0202239}{[arXiv:hep-ph/0202239 [hep-ph]]}.

\bibitem{Watson:2009hw}
S.~Watson,
``Reevaluating the Cosmological Origin of Dark Matter,''
Adv. Ser. Direct. High Energy Phys. \textbf{21} (2010), 305-324
doi:10.1142/9789814307505\_0007
\href{https://arxiv.org/abs/0912.3003}{[arXiv:0912.3003 [hep-th]]}.

\bibitem{Acharya:2008bk}
B.~S.~Acharya, P.~Kumar, K.~Bobkov, G.~Kane, J.~Shao and S.~Watson,
``Non-thermal Dark Matter and the Moduli Problem in String Frameworks,''
JHEP \textbf{06} (2008), 064
\href{https://iopscience.iop.org/article/10.1088/1126-6708/2008/06/064}{doi:10.1088/1126-6708/2008/06/064}
\href{https://arxiv.org/abs/0804.0863}{[arXiv:0804.0863 [hep-ph]]}.

\bibitem{Acharya:2009zt}
B.~S.~Acharya, G.~Kane, S.~Watson and P.~Kumar,
``A Non-thermal WIMP Miracle,''
Phys. Rev. D \textbf{80} (2009), 083529
\href{https://journals.aps.org/prd/abstract/10.1103/PhysRevD.80.083529}{doi:10.1103/PhysRevD.80.083529}
\href{https://arxiv.org/abs/0908.2430}{[arXiv:0908.2430 [astro-ph.CO]]}.

\bibitem{Kofman:1997yn}
L.~Kofman, A.~D.~Linde and A.~A.~Starobinsky,
``Towards the theory of reheating after inflation,''
Phys. Rev. D \textbf{56} (1997), 3258-3295
\href{https://journals.aps.org/prd/abstract/10.1103/PhysRevD.56.3258}{doi:10.1103/PhysRevD.56.3258}
\href{https://arxiv.org/abs/hep-ph/9704452}{[arXiv:hep-ph/9704452 [hep-ph]]}.

\bibitem{Kofman:1994rk}
L.~Kofman, A.~D.~Linde and A.~A.~Starobinsky,
``Reheating after inflation,''
\href{https://journals.aps.org/prl/abstract/10.1103/PhysRevLett.73.3195}{Phys. Rev. Lett. \textbf{73} (1994), 3195-3198}
doi:10.1103/PhysRevLett.73.3195
\href{https://arxiv.org/abs/hep-th/9405187}{[arXiv:hep-th/9405187 [hep-th]]}.

\bibitem{Shtanov:1994ce}
Y.~Shtanov, J.~H.~Traschen and R.~H.~Brandenberger,
``Universe reheating after inflation,''
Phys. Rev. D \textbf{51} (1995), 5438-5455
\href{https://journals.aps.org/prd/abstract/10.1103/PhysRevD.51.5438}{doi:10.1103/PhysRevD.51.5438}
\href{https://arxiv.org/abs/hep-ph/9407247}{[arXiv:hep-ph/9407247 [hep-ph]]}.

\bibitem{Starobinsky:1980te}
A.~A.~Starobinsky,
``A New Type of Isotropic Cosmological Models Without Singularity,''
Phys. Lett. B \textbf{91} (1980), 99-102
\href{https://www.sciencedirect.com/science/article/abs/pii/037026938090670X?via%3Dihub}{doi:10.1016/0370-2693(80)90670-X}

\bibitem{Guth:1980zm}
A.~H.~Guth,
``The Inflationary Universe: A Possible Solution to the Horizon and Flatness Problems,''
Phys. Rev. D \textbf{23} (1981), 347-356
\href{https://journals.aps.org/prd/abstract/10.1103/PhysRevD.23.347}{doi:10.1103/PhysRevD.23.347}

\bibitem{Linde:1981mu}
A.~D.~Linde,
``A New Inflationary Universe Scenario: A Possible Solution of the Horizon, Flatness, Homogeneity, Isotropy and Primordial Monopole Problems,''
Phys. Lett. B \textbf{108} (1982), 389-393
\href{https://www.sciencedirect.com/science/article/abs/pii/0370269382912199?via%3Dihub}{doi:10.1016/0370-2693(82)91219-9}

\bibitem{Bodeker:2020ghk}
D.~B\"odeker and W.~Buchm\"uller,
``Baryogenesis from the weak scale to the grand unification scale,''
\href{https://arxiv.org/abs/2009.07294}{[arXiv:2009.07294 [hep-ph]]}.

\bibitem{Greene:1997fu}
P.~B.~Greene, L.~Kofman, A.~D.~Linde and A.~A.~Starobinsky,
``Structure of resonance in preheating after inflation,''
Phys. Rev. D \textbf{56} (1997), 6175-6192
\href{https://journals.aps.org/prd/abstract/10.1103/PhysRevD.56.6175}{doi:10.1103/PhysRevD.56.6175}
\href{https://arxiv.org/abs/hep-ph/9705347}{[arXiv:hep-ph/9705347 [hep-ph]]}.

\bibitem{Felder:1998vq}
G.~N.~Felder, L.~Kofman and A.~D.~Linde,
``Instant preheating,''
Phys. Rev. D \textbf{59} (1999), 123523
\href{https://journals.aps.org/prd/abstract/10.1103/PhysRevD.59.123523}{doi:10.1103/PhysRevD.59.123523}
\href{https://arxiv.org/abs/hep-ph/9812289}{[arXiv:hep-ph/9812289 [hep-ph]]}.

\bibitem{Podolsky:2005bw}
D.~I.~Podolsky, G.~N.~Felder, L.~Kofman and M.~Peloso,
``Equation of state and beginning of thermalization after preheating,''
Phys. Rev. D \textbf{73} (2006), 023501
\href{https://journals.aps.org/prd/abstract/10.1103/PhysRevD.73.023501}{doi:10.1103/PhysRevD.73.023501}
\href{https://arxiv.org/abs/hep-ph/0507096}{[arXiv:hep-ph/0507096 [hep-ph]]}.

\bibitem{Davidson:2000er}
S.~Davidson and S.~Sarkar,
``Thermalization after inflation,''
JHEP \textbf{11} (2000), 012
doi:10.1088/1126-6708/2000/11/012
\href{https://arxiv.org/abs/hep-ph/0009078}{[arXiv:hep-ph/0009078 [hep-ph]].}

\bibitem{Harigaya:2013vwa}
K.~Harigaya and K.~Mukaida,
``Thermalization after/during Reheating,''
JHEP \textbf{05} (2014), 006
\href{https://link.springer.com/article/10.1007/JHEP05(2014)006}{doi:10.1007/JHEP05(2014)006}
\href{https://arxiv.org/abs/1312.3097}{[arXiv:1312.3097 [hep-ph]]}.

\bibitem{Drewes:2013iaa}
M.~Drewes and J.~U.~Kang,
``The Kinematics of Cosmic Reheating,''
Nucl. Phys. B \textbf{875} (2013), 315-350
[erratum: Nucl. Phys. B \textbf{888} (2014), 284-286]
\href{https://www.sciencedirect.com/science/article/pii/S0550321313003751?via%3Dihub}{doi:10.1016/j.nuclphysb.2013.07.009}
\href{https://arxiv.org/abs/1305.0267}{[arXiv:1305.0267 [hep-ph]]}.

\bibitem{Drewes:2015eoa}
M.~Drewes and J.~U.~Kang,
``Sterile neutrino Dark Matter production from scalar decay in a thermal bath,''
JHEP \textbf{05} (2016), 051
\href{https://link.springer.com/article/10.1007/JHEP05(2016)051}{doi:10.1007/JHEP05(2016)051}
\href{https://arxiv.org/abs/1510.05646}{[arXiv:1510.05646 [hep-ph]]}.

\bibitem{Mukaida:2012bz}
K.~Mukaida and K.~Nakayama,
``Dissipative Effects on Reheating after Inflation,''
JCAP \textbf{03} (2013), 002
doi:10.1088/1475-7516/2013/03/002
\href{https://arxiv.org/abs/1212.4985}{[arXiv:1212.4985 [hep-ph]]}.

\bibitem{Giudice:2000ex}
G.~F.~Giudice, E.~W.~Kolb and A.~Riotto,
``Largest temperature of the radiation era and its cosmological implications,''
Phys. Rev. D \textbf{64} (2001), 023508
doi:10.1103/PhysRevD.64.023508
[arXiv:hep-ph/0005123 [hep-ph]].

\bibitem{Drewes:2014pfa}
M.~Drewes,
``On finite density effects on cosmic reheating and moduli decay and implications for Dark Matter production,''
JCAP \textbf{11} (2014), 020
\href{https://iopscience.iop.org/article/10.1088/1475-7516/2014/11/020}{doi:10.1088/1475-7516/2014/11/020}
\href{https://arxiv.org/abs/1406.6243}{[arXiv:1406.6243 [hep-ph]]}.

\bibitem{Drewes:2015coa}
M.~Drewes,
``What can the CMB tell about the microphysics of cosmic reheating?,''
JCAP \textbf{03} (2016), 013
\href{https://iopscience.iop.org/article/10.1088/1475-7516/2016/03/013}{doi:10.1088/1475-7516/2016/03/013}
\href{https://arxiv.org/abs/1511.03280}{[arXiv:1511.03280 [astro-ph.CO]]}.

\bibitem{Co:2020xaf}
R.~T.~Co, E.~Gonzalez and K.~Harigaya,
``Increasing Temperature toward the Completion of Reheating,''
JCAP \textbf{11} (2020), 038
\href{https://iopscience.iop.org/article/10.1088/1475-7516/2020/11/038}{doi:10.1088/1475-7516/2020/11/038}
\href{https://arxiv.org/abs/2007.04328v1}{[arXiv:2007.04328 [astro-ph.CO]]}.

\bibitem{Garcia:2020wiy}
M.~A.~G.~Garcia, K.~Kaneta, Y.~Mambrini and K.~A.~Olive,
``Inflaton Oscillations and Post-Inflationary Reheating,''
JCAP \textbf{04} (2021), 012
\href{https://arxiv.org/abs/2012.10756}{[arXiv:2012.10756 [hep-ph]]}.

\bibitem{Yokoyama:2006wt}
J.~Yokoyama,
``Thermal background can solve the cosmological moduli problem,''
Phys. Rev. Lett. \textbf{96} (2006), 171301
\href{https://journals.aps.org/prl/abstract/10.1103/PhysRevLett.96.171301}{doi:10.1103/PhysRevLett.96.171301}
\href{https://arxiv.org/abs/hep-ph/0601067}{[arXiv:hep-ph/0601067 [hep-ph]]}.

\bibitem{Bodeker:2006ij}
D.~Bodeker,
``Moduli decay in the hot early Universe,''
JCAP \textbf{06} (2006), 027
\href{https://iopscience.iop.org/article/10.1088/1475-7516/2006/06/027}{doi:10.1088/1475-7516/2006/06/027}
\href{https://arxiv.org/abs/hep-ph/0605030}{[arXiv:hep-ph/0605030 [hep-ph]]}.

\bibitem{Ming:2019qty}
H.~Xu, L.~Ming and Y.~K.~E.~Cheung,
``Dynamics of scalar fields in an expanding/contracting cosmos at finite temperature,''
Chin. Phys. C \textbf{44} (2020) no.5, 053103
\href{https://iopscience.iop.org/article/10.1088/1674-1137/44/5/053103}{doi:10.1088/1674-1137/44/5/053103}
\href{https://arxiv.org/abs/1904.12941}{[arXiv:1904.12941 [hep-th]]}.

\bibitem{Carenza:2019vzg}
P.~Carenza, A.~Mirizzi and G.~Sigl,
``Dynamical evolution of axion condensates under stimulated decays into photons,''
Phys. Rev. D \textbf{101} (2020) no.10, 103016
\href{https://journals.aps.org/prd/abstract/10.1103/PhysRevD.101.103016}{doi:10.1103/PhysRevD.101.103016}
\href{https://arxiv.org/abs/1911.07838}{[arXiv:1911.07838 [hep-ph]]}.

\bibitem{Drewes:2019rxn}
M.~Drewes,
``Measuring the Inflaton Coupling in the CMB,''
\href{https://arxiv.org/abs/1903.09599}{[arXiv:1903.09599 [astro-ph.CO]]}.
\end{thebibliography}
\end{document}